# Laser Noise Reduction in Air

Pierre Béjot
*GAP, Université de Genève, 20 rue de l'École de médecine, CH 1211Genève, Switzerland*
Jérôme Kasparian[*], Estelle Salmon, Roland Ackermann
*LASIM, UMR CNRS 5579, Université Lyon 1, 43 bd du 11 Novembre 1918, F69622 Villeurbanne Cedex, France*
Nicolas Gisin, Jean-Pierre Wolf
*GAP, Université de Genève, 20 rue de l'École de médecine, CH 1211Genève, Switzerland*

## *Abstract*

Fluctuations of the white-light supercontinuum produced by ultrashort laser pulses in self-guided filaments (spatio-temporal solitons) in air are investigated. We demonstrate that correlations exist within the white-light supercontinuum, and that they can be used to significantly reduce the laser intensity noise by filtering the spectrum. More precisely, the fundamental wavelength is anticorrelated with the wings of the continuum, while conjugated wavelength pairs on both sides of the continuum are strongly correlated. Spectral filtering of the continuum reduces the laser intensity noise by 1.2 dB, showing that fluctuations are rejected to the edges of the spectrum.

Considerable interest has recently been devoted to quantum optics and non-linear effects in transparent media. Second order nonlinearity ($\chi^2$) processes in parametric generators have been the model of choice in this respect, where both photon correlation and squeezing were first demonstrated. Recent studies showed that both phenomena also occurred for temporal solitons in optical fibers [1,2]. The origin of the correlations in the generated continuum is intrinsic to the ($\chi^3$) self-phase-modulation process [3]. For sufficient laser intensity, self-organized spatio-temporal solitons form even in the air. These self-generated light filaments propagate over several hundreds of meters and give rise to an exceptionally broad continuum [4]. In this Letter, we report the observation of both spectral correlations in the continuum and laser noise reduction for filaments propagating in air. These results, which remain in the classical domain of correlations and noise reduction, open perspectives for high precision remote measurements of atmospheric molecules or for the transmission of encrypted information [5]. For instance, water vapour concentration and atmospheric temperature profile measurements require a precision better than 1% to be useful for global warming models. A significant source of noise in Lidar measurements, besides atmospheric fluctuations, is laser noise. A low noise broadband laser covering several absorption bands of $H_2O$ would therefore be an ideal source for such measurements (the carrier wavelength would then be slightly shifted to 830 nm).

Filaments [6] arise in the non-linear propagation of ultrashort, high-power laser pulses in transparent media. They result from a subtle balance between Kerr-lens focusing and defocusing by self-induced plasma. In the atmosphere, filaments have been observed over several hundreds of meters, up to a few kilometers away from the laser source [7], even in perturbed conditions such as clouds [8] or turbulence [9]. These properties open the way to atmospheric applications [4], such as Lidar (Light Detection and Ranging) remote sensing, Laser-induced breakdown spectroscopy (LIBS), lightning control or free space communications. Up to now, filamentation has been studied extensively, but only from the classical point of view. Quantum optics of both self guiding and white-light generation by self-phase modulation (SPM) and the resulting correlations in the spectrum of the white-light continuum have not been described so far. Here, we demonstrate that correlations exist within the white-light supercontinuum, and that they can be used to significantly reduce the laser intensity noise by filtering the spectrum. These result on non-linear optics in the air show that, even at very high intensities, the broadened spectrum is the result of interactions of the type $\omega_1+\omega_2 = \omega_3+\omega_4$, origin of the observed correlations. In particular, in the first phase of the spectral broadening process, the most intensity is at $\omega_0$, and we expect correlations for $2\omega_0 = \omega_1 + \omega_2$

The experimental setup is depicted in Figure 1. A CPA (Chirped Pulse Amplification) Ti:Sapphire laser system delivered 200 fs pulses of typ. 1 mJ, at 22.5 Hz repetition rate. The central wavelength of the laser was set at 805 and 817 nm depending on the alignment, with an initial bandwidth of 8 nm. The beam was focused by a spherical mirror of 5 m focal length, yielding a non-linear focus (filament onset) ~ 3 m downstream of the spherical mirror: this non-linear focus was defined as origin of the propagation axis ($z = 0$). The filament length was ~ 4 m. At $z = 10$ m, the continuum generated by the filament was dispersed by a diffraction grating (order 1 blazed, efficiency 0.9 at 810 nm), and two photodiodes (quantum efficiency: 0.93 at 810nm) selected two specific spectral channels (9 nm bandwidth). The correlation coefficient between these two wavelength channels was numerically evaluated for each time series (5000 shots). Reference conditions without non-linear propagation were obtained by using low energy and a plane mirror instead of the spherical one in order to avoid filamentation.

In a second experiment, the filament was scattered on an achromatic target, and analyzed by a high resolution spectrometer (0.3 nm) between 785 and 845 nm. 5000 spectra were recorded, normalized to unity and used to compute the cross-correlation map across the spectrum. In a third experiment, we directly measured the laser intensity noise reduction factor. For this, we used the "low loss" grating setup and a single photodiode recorded a 20 000-shots time series of the integrated spectrum over different spectral intervals. The standard deviation, for both focused (filamenting) and unfocused (reference) conditions was computed to quantify the noise reduction due to the SPM correlations within the filament.

Figure 2 displays the cross-correlation of the white-light continuum with one fixed wavelength at $\lambda_2 = 844$ nm as measured with the setup of Figure 1. The central wavelength of the fundamental spectrum was $\lambda_0 = 805$ nm in this experiment. As expected from the quantum point of view, where two photons at the fundamental wavelength $\lambda_0$ are converted into a pair of photons at wavelengths $\lambda_1$ and $\lambda_2$, with $2/\lambda_0 = 1/\lambda_1 + 1/\lambda_2$, a strong correlation ($C_{\lambda_1 \lambda_2} = 0.85$) is observed at the conjugate wavelength of the fixed wavelength. In contrast, the fundamental, which is depleted when the white light is generated, is anticorrelated ($C_{\lambda_0 \lambda_1} = -0.4$). The typical number of photons measured in the experiment is $10^8$ per shot. In order to determine whether the correlations are classical or quantum, we evaluated the "Gemellity Factor" [10]: $G = \frac{F_1 + F_2}{2} - \sqrt{C_{12}^2 F_1 F_2 + \frac{(F_1 - F_2)^2}{4}}$ where $F_1$ and $F_2$ are the Fano factors of the two wavelengths, *i.e.* the ratio of the signal noise at the considered wavelength, over the shot noise [11]. G amounts to $10^7$, i.e. much greater than unity, for any wavelength pair within the 760-844 nm range. Hence, although the number of detected photons is moderate and the losses in this first experiment are limited, the observed correlations are classical and represent the correlation between the instantaneous fluctuations of the different photon fluxes.

The occurrence of correlations within the spectrum was further confirmed by considering the whole map of cross-correlations (Figure 3), as obtained from the second setup, which involves high losses. In this experiment, $\lambda_0 = 815$ nm. Positive correlations are observed in regions corresponding to nearly conjugated wavelengths, as well as in the trivial case of $\lambda_1 = \lambda_2$ axis. In contrast, negative correlations form a dark cross centered on the fundamental wavelength. In other words, the generation of the white-light photons requires a depletion of the fundamental photon number. It is interesting to notice that structures (lines) appear in the map, both in the positive and in the negative correlation regions. This behavior can be interpreted as cascades of 4 photons mixing processes, where one (or both) of the conjugated wavelengths is (are) involved in a consecutive process, destroying the initial correlations. Moreover, the probability of cascading depends on the spectral intensity, which is related to the broadening itself.

Spectral filtering the supercontinuum around $\lambda_0 = 805$ nm (785-820 nm) and comparing the intensity fluctuations between the filamenting and the reference (linearly propagating) beams (setup #3) demonstrated a reduction of the intensity noise power (*i.e.* the variance of the recorded photocurrent) by 1.2 dB. In contrast, a spectral range restricted to one side of the fundamental wavelength $\lambda_0$ (814-849 nm) yielded no noise reduction. As observed in the case of fibers, SPM rejects the fluctuations to the edges of the spectrum.

Considering the very high intensity involved to generate filaments in air, these results are particularly unexpected. In particular, it is remarkable that the presence of the self-induced plasma, which stabilizes the filamentation process, has no significant effect on the correlations in the spectrum.

We have demonstrated that the broadband continuum generated by spatio-temporal solitons propagating in air (filaments) exhibit strong correlations. Although obtained at very high intensities, these correlations are the scars of quantum 4-photons interactions. As fluctuations are rejected to the edges of the spectrum, spectral filtering of the continuum allowed a reduction of the laser intensity noise by 1.2 dB, still well above shot noise. Moreover, the characteristics of the spectral filter were not optimized in this demonstration experiment, and higher compression ratios can be expected. These results might open a new range of applications, such as high precision Lidar measurements of atmospheric pollutants or transmission schemes in optical communications.

*Acknowledgements.* We gratefully acknowledge H. Zbinden (Université de Genève), C. Fabre (ENS, Paris), and G. Leuchs (Universitaet Erlangen) for very fruitful discussions.

## *Figures*

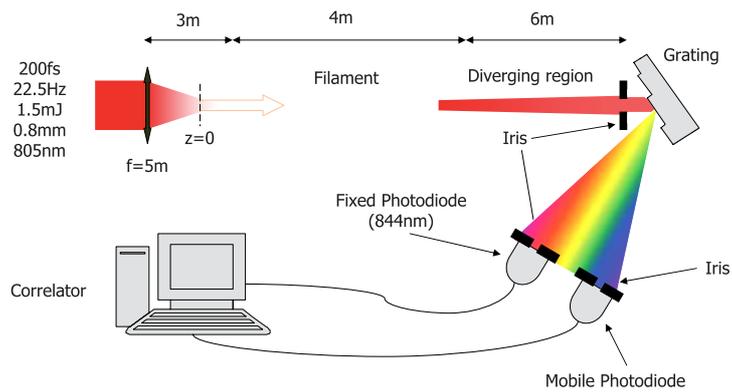

Figure 1. Low loss experimental setup used for the measurement of correlations within the light supercontinuum generated by filaments

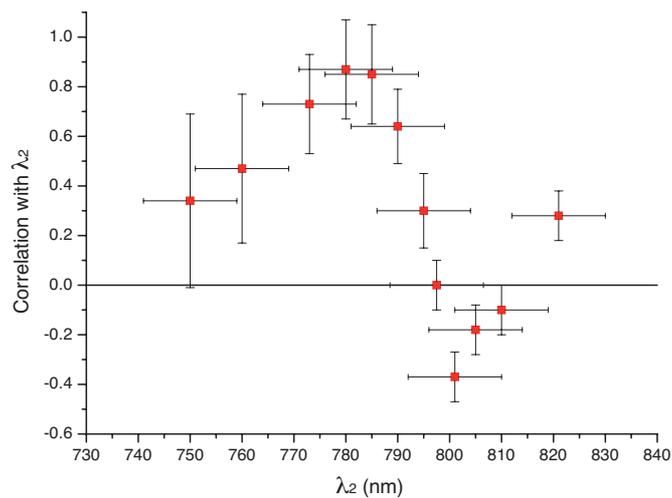

Figure 2. Demonstration of correlations within the continuum spectrum. Strong correlations are observed for the wavelengths $\lambda_2$ conjugated to the fixed wavelength channel about $\lambda_1 =$ 840 nm, while anticorrelation is observed with the fundamental spectrum centered on $\lambda_0 =$ 805 nm.

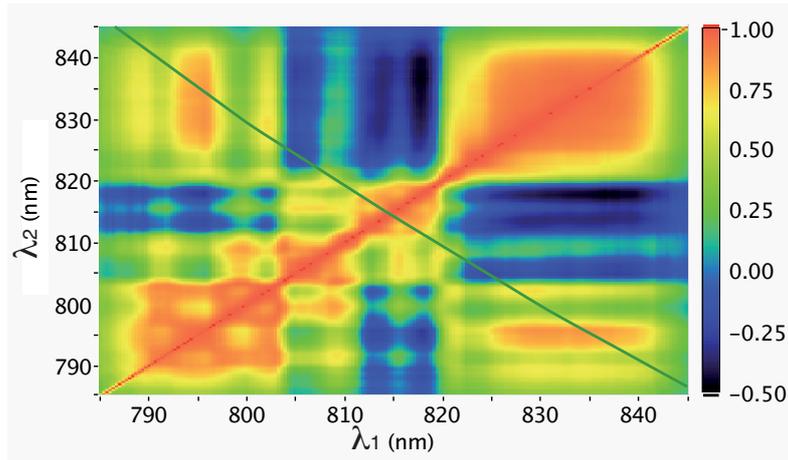

Figure 3. 2D-Map of spectral correlations within the filament continuum. The green line corresponds to the wavelength pairs that satisfy the relation $2/\lambda_0 = 1/\lambda_1 + 1/\lambda_2$ with $\lambda_0 = 815$nm.